\documentclass[doublecol]{epl2}
\usepackage{bm}
\usepackage{amsmath}
\usepackage{amssymb}

\begin{document}

\title{A three-sphere microswimmer in a structured fluid}
\shorttitle{A three-sphere microswimmer in a structured fluid}

\author{Kento Yasuda,\inst{1} Ryuichi Okamoto \inst{2} and Shigeyuki Komura \inst{1}}
\shortauthor{K. Yasuda, R. Okamoto and S. Komura}

\institute{
Department of Chemistry, Graduate School of Science,
Tokyo Metropolitan University, Tokyo 192-0397, Japan 
}
\institute{
\inst{1} Department of Chemistry, Graduate School of Science,
Tokyo Metropolitan University, Tokyo 192-0397, Japan \\
\inst{2} Research Institute for Interdisciplinary Science, Okayama University,
Okayama 700-8530, Japan
}


\abstract{
We discuss the locomotion of a three-sphere microswimmer in a viscoelastic structured fluid   
characterized by typical length and time scales. 
We derive a general expression to link the average swimming velocity to the sphere mobilities. 
In this relationship, a viscous contribution exists when the time-reversal symmetry is broken, whereas 
an elastic contribution is present when the structural symmetry of the microswimmer is broken.
As an example of a structured fluid, we consider a polymer gel, which is described by a ``two-fluid" model.
We demonstrate in detail that the competition between the swimmer size and the polymer mesh
size gives rise to the rich dynamics of a three-sphere microswimmer.
}

\maketitle

\section{Introduction}

Microswimmers are tiny machines, such as sperm cells or motile bacteria, that swim in a fluid
and are expected to be relevant to microfluidics and microsystems~\cite{Lauga09a}.
By transforming chemical energy into mechanical work, microswimmers can change their shapes and 
move in viscous environments.
The fluid forces acting on the length scale of microswimmers are governed by the effect of viscous 
dissipation. 
According to Purcell's scallop theorem~\cite{Purcell77}, time-reversal body motion cannot be used 
for locomotion in a Newtonian fluid.
As one of the simplest models exhibiting broken time-reversal symmetry, Najafi and Golestanian 
proposed a three-sphere microswimmer~\cite{Golestanian04,Golestanian08} in which three in-line 
spheres are linked by two arms of varying lengths. 
This model is suitable for analytical studies because the tensorial structure of the fluid motion can 
be neglected in its translational motion.
Recently, such a microswimmer has been experimentally realized~\cite{Grosjean16,Grosjean18}.

For microswimmers in general situations, however, the surrounding fluid is not necessarily purely 
viscous but viscoelastic.
Several studies have discussed the swimming behaviors of micromachines in different types of viscoelastic 
fluids~\cite{Lauga09b,Teran10,Curtis13,Qiu14,Ishimoto17}.
In particular, Lauga showed that the Scallop theorem in a viscoelastic fluid breaks down if the 
squirmer has a fore-aft asymmetry in its surface velocity distribution~\cite{Lauga09b}.
In a recent study, we discussed the locomotion of a three-sphere microswimmer in a 
viscoelastic medium and derived a relationship linking the average swimming velocity to 
the frequency-dependent viscosity of the surrounding medium~\cite{Yasuda17a}.
We demonstrated that the absence of the time-reversal symmetry of the swimmer motion is reflected 
in the real part of the viscosity, whereas the absence of the structural symmetry of the swimmer is 
reflected in its imaginary part.

So far, investigations into the swimming behaviors of micromachines have been limited to 
homogeneous viscoelastic fluids without any internal structures. 
However, one of the fundamental and characteristic features of viscoelastic soft matter is that it 
contains various intermediate mesoscopic structures and behaves as a structured 
fluid~\cite{WittenBook}.
The existence of such internal length scales significantly affects the rheological properties of 
soft matter~\cite{LarsonBook}. 
In this letter, we address the effects of the intermediate structures of the surrounding 
viscoelastic fluid on the locomotion of a three-sphere microswimmer.
Because a three-sphere microswimmer is also characterized by its own size, our main interest is to 
find out how the average swimming velocity depends on the relative magnitudes of the 
swimmer's size and the characteristic length of the surrounding fluid.

\begin{figure}[bth]
\begin{center}
\includegraphics[scale=0.5]{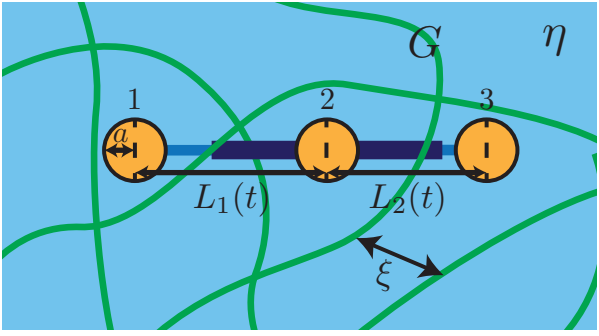}
\end{center}
\caption{
A three-sphere micromachine swimming in a structured fluid such as a polymer gel. 
Three identical spheres of radius $a$ are connected by arms with lengths $L_1(t)$ and $L_2(t)$
and undergo time-dependent cyclic motions. 
According to the two-fluid model, the polymer gel consists of an elastic network characterized 
by a constant shear modulus $G$ and a viscous fluid characterized by a constant shear 
viscosity $\eta$. 
The elastic and fluid components are coupled via mutual friction. 
The length scale $\xi$ characterizes the typical internal structure of the elastic network, e.g., its mesh size.
}
\label{model}
\end{figure}

Generalizing our previous work~\cite{Yasuda17a}, we first obtain the average velocity of a 
three-sphere microswimmer moving in a structured fluid, which is characterized by typical length 
and time scales.
As an example of a structured fluid, we employ a ``two-fluid" model that has been broadly 
used to describe the dynamics of polymer gels~\cite{deGennes76a,deGennes76b,Brochard77}.
Recently, the response of a polymer network to the motion of a rigid sphere has been investigated 
within this two-fluid model~\cite{Diamant15,Sonn14,Sonn14-b}.
We calculate the frequency dependency of the average velocity of a three-sphere microswimmer in a 
two-fluid gel and obtain its various asymptotic expressions by changing the swimmer size.
The competition between the swimmer size and the polymer mesh size gives rise to 
the rich dynamics of microswimmers. 
Even though we primarily discuss the two-fluid model here, our result can be applied to various 
types of structured fluids.

\section{Microswimmer in a structured fluid}

As shown in fig.~\ref{model}, we consider a microswimmer consisting of three rigid spheres 
of the same radius $a$ that are connected by two arms of variable lengths $L_1$ and 
$L_2$~\cite{Golestanian04,Golestanian08}. 
We assume that the motion of the arms is prescribed by two time-dependent functions 
$L_1(t), L_2(t) \gg a$.
Then the velocity of each sphere $V_i$ ($i=1,2,3$) should  satisfy the conditions 
$\dot L_1(t)=V_2(t)-V_1(t)$ and $\dot L_2(t)=V_3(t)-V_2(t)$, 
where the dot indicates the time derivative.
The surrounding fluid exerts a force $F_i$ ($i=1,2,3$)
on each sphere, which we assume to be along the swimmer axis. 
Because we are interested in the autonomous net locomotion of the swimmer, there are no external 
forces acting on the spheres. 
This leads to the force-free condition: $F_1(t)+F_2(t)+F_3(t)=0$.

Within the linear response theory, the velocity and force acting on a sphere of radius $a$ 
are related in the Fourier domain by $V_i(\omega)=\mu[a,\omega]F_i(\omega)$, 
where $V(\omega) =\int_{-\infty}^{\infty} dt \, V(t)e^{-i\omega t}$ 
(with the same form for $F(\omega)$) denotes the Fourier 
transform and $\mu[a,\omega] =\int_{0}^{\infty} dt  \, \mu(a,t)e^{-i\omega t}$ 
gives the frequency-dependent self-mobility.
Similarly, the force $F_{j}$ acting on the $j$-th sphere at $x_j$ and the induced velocity $V_i$ 
of the $i$-th sphere at $x_i$ are related by $V_i(\omega)=M[r,\omega]F_{j}(\omega)$, 
where $r=x_i-x_j \gg a$ and $M[r,\omega]$ is the frequency-dependent longitudinal coupling mobility.

We further assume that the arm deformations are relatively small, and given by 
$L_1(t)=\ell+u_1(t)$ and $L_2(t)=\ell+u_2(t)$, where $\ell$ is a constant length that satisfies 
$\ell \gg u_1(t), u_2(t)$.
We consider the case when the two arms undergo the simplest periodic 
motions~\cite{Golestanian04,Golestanian08}:
$u_1(t)=d_1\cos(\Omega t)$ and $u_2(t)=d_2\cos(\Omega t-\phi)$,
where $d_1$ and $d_2$ are the amplitudes of the oscillatory motions, $\Omega$ is the 
common arm frequency, and $\phi$ is the mismatch in the phases between the two arms.
When the arm motions are given, the above set of equations is sufficient to solve for the six unknown 
quantities $V_i$ and $F_i$.
The swimming velocity is obtained by averaging the velocities of the three spheres, i.e., 
$V=(V_1+ V_2+ V_3)/3$.

Consider a viscoelastic structured fluid that is characterized by a characteristic length
scale $\xi$ and a characteristic time scale $\tau$. 
We assume that the above mentioned mobilities are expressed by the following scaling forms:
\begin{align}
\mu[a,\omega]=\frac{\hat \mu[a/\xi,\omega\tau]}{6\pi\eta_0a},~~~~~
M[r,\omega]=\frac{\hat M[r/\xi,\omega\tau]}{4\pi\eta_0\ell},
\label{scaling}
\end{align}
where $\hat \mu$ and $\hat M$ are the dimensionless scaling functions and $\eta_0$ 
is the zero-frequency shear viscosity.
Even if there are more than two length or time scales, the above assumption is still valid 
because only the dimensionless ratios between the different scales enter into the scaling functions. 
In other words, if there are several length scales $\xi_1$, $\xi_2$, $\xi_3, \cdots$ and several 
time scales $\tau_1$, $\tau_2$, $\tau_3, \cdots$, the dimensionless mobility can be expressed as 
$\hat \mu[a/\xi_1,\omega\tau_1; \xi_2/\xi_1, \xi_3/\xi_1, \cdots; \tau_2/\tau_1, 
\tau_3/\tau_1, \cdots]$ and similarly for $\hat M$.
Under this assumption, we perform an expansion of the swimming velocity to the leading order in 
$a/\ell$, $d_1/\ell$, and $d_2/\ell$.
After performing the time integration over a full cycle, we obtain the average swimming velocity: 
\begin{widetext}
\begin{align}
\overline V & \approx\frac{d_1d_2a\Omega}{48\ell^2}\left(\hat\mu[a/\xi,\Omega\tau]^{-1}
+\hat\mu[a/\xi,-\Omega\tau]^{-1}\right)\left(8\hat M[\ell/\xi,0]-\hat M[2\ell/\xi,0]
+2\ell(-4\partial \hat M[\ell/\xi,0]+\partial \hat M[2\ell/\xi,0])\right)\sin\phi\nonumber\\
&+\frac{i(d_1^2-d_2^2)a\Omega}{96\ell^2}\left(\hat\mu[a/\xi,\Omega\tau]^{-1}
-\hat\mu[a/\xi,-\Omega \tau]^{-1}\right)
\left(4\hat M[\ell/\xi,0]+\hat M[2\ell/\xi,0]
-2\ell(2\partial \hat M[\ell/\xi,0]+\partial \hat M[2\ell/\xi,0])\right),
\label{barVve}
\end{align}
\end{widetext}
\begin{equation}
\mbox{\textit{see eq.~\eqref{barVve}}}
\nonumber
\end{equation}
where 
$\partial \hat M[\ell/\xi,0]=\lim_{\omega \tau \to 0} (\partial \hat M[r/\xi,\omega\tau]/\partial r)_{r=\ell}$
(see the SM for the full derivation).
This is a generalization of our previous result~\cite{Yasuda17a} and is the main result of this letter.

The first term in eq.~(\ref{barVve}) can be regarded as a viscous contribution, $\overline V_{\rm v}$, 
and is present only if the time-reversal symmetry of the arm motion is broken, i.e., 
$\phi \ne 0,\pi$.
The second term, conversely, corresponds to an elastic contribution, $\overline V_{\rm e}$, 
and exists only when the structural symmetry of the swimmer is broken, i.e., $d_1 \ne d_2$. 
In other words, even if the time-reversal symmetry of the swimmer motion is not broken, i.e., 
$\phi= 0,\pi$, the swimmer can still move in a viscoelastic medium because of 
the second elastic term as long as its structural symmetry is broken, i.e., $d_1\ne d_2$. 
We used the condition $a \ll \ell$ when deriving eq.~(\ref{barVve}), but nothing has been 
assumed concerning the relative magnitudes between the swimmer size, $a$ and $\ell$, 
and the characteristic length of the fluid, $\xi$.
Therefore, eq.~(\ref{barVve}) offers a very general velocity expression for a three-sphere microswimmer
moving in a structured fluid.

Although eq.~(\ref{barVve}) is applicable to any structured fluid, some special cases are worth 
discussing.
For a purely viscous fluid, the scaling functions in eq.~(\ref{scaling}) are given by 
$\hat\mu=1$ and $\hat M=1$; therefore, eq.~(\ref{barVve})
reduces to the average velocity obtained by Golestanian and Ajdari~\cite{Golestanian08}.
Further, for a viscoelastic fluid without any internal structure, the scaling functions 
are simply given by $\hat \mu=\eta_0/\eta[\omega]$ and 
$\hat M=\eta_0/\eta[\omega]$, where $\eta[\omega]$ is the 
frequency-dependent complex viscosity.
In such a homogeneous but viscoelastic fluid, eq.~(\ref{barVve}) reduces to eq.~(13) in 
ref.~\cite{Yasuda17a}.

We note that the above derivation has been limited within the linear response 
theory because linear relationships between forces and velocities have been assumed.
Such an assumption of the linear viscoelasticity is generally justified when the strain amplitude
is small enough~\cite{PipkinBook}.
For a three-sphere microswimmer, this condition is given by $d_1/\ell \ll 1$ and $d_2/\ell \ll 1$
which have been indeed used in the derivation of eq.~(\ref{barVve}). 
Otherwise, one needs to take into account nonlinear viscoelastic effects such as a shear thinning
behavior~\cite{LarsonBook}.

\section{Two-fluid model for a gel}

As a simple example of various structured fluids, we consider here a polymer gel described by 
the two-fluid model. 
As schematically described in fig.~\ref{model}, there are two dynamical fields in this 
model: the displacement field $\mathbf{u}(\mathbf{r},t)$ 
of the elastic network and the velocity field $\mathbf{v}(\mathbf{r},t)$ of the permeating fluid.
When inertial effects are neglected, the linearized coupled equations for these two field variables 
are given by 
\begin{align}
0
&=G \nabla^2 \mathbf{u}+(K+G/3)\nabla(\nabla\cdot \mathbf{u})-\Gamma 
\left(\frac{\partial \mathbf{u}}{\partial t} -\mathbf{v} \right),
\label{TF-model1}
\end{align}
\begin{align}
0=
\eta \nabla^2 \mathbf{v} -\nabla p 
-\Gamma \left(\mathbf{v}- \frac{\partial \mathbf{u}}{\partial t}\right)+\mathbf{f}.
\label{TF-model2}
\end{align}
Here, $G$ and $K$ are the shear and compression moduli of the elastic network, respectively,
$\eta$ is the shear viscosity of the fluid,  $p(\mathbf{r},t)$ is the pressure field, and 
$\mathbf{f}(\mathbf{r},t)$ is the external force density acting on the fluid component.
The elastic and fluid components are coupled via the mutual friction, which is characterized 
by the friction coefficient $\Gamma$. 
When the volume fraction of the elastic component is small, we further require the incompressibility
condition: $\nabla \cdot \mathbf{v} = 0$.
The above two-fluid model contains the characteristic length $\xi=(\eta/\Gamma)^{1/2}$ 
and the characteristic time $\tau=\eta/G$.
The former length scale roughly corresponds to the mesh size of a polymer network, and the latter 
time scale sets the viscoelastic time.
Hereafter, we introduce a dimensionless ratio defined as $\epsilon =[(K+4G/3)/G]^{1/2}$.

Diamant calculated the self-mobility of a sphere in a two-fluid gel, which depends on the choice of the 
boundary condition at the surface of the sphere~\cite{Diamant15}.
Here we consider the case of ``a sticking fluid and a free network", i.e., a stick boundary condition 
is used for the fluid while the network does not exchange stress with the sphere.
In other words, the network moves only because of its coupling to the fluid.
The full expression of the self-mobility $\mu$ is given in the SM~\cite{Diamant15}, 
and here, we show only its limiting behaviors.
In the low-frequency limit, $\omega \tau \to 0$, $\mu$ becomes 
\begin{align}
&\mu[a,\omega]  \approx \frac{9}{6\pi\eta a (9+9\hat a+\hat a^2)}
\nonumber \\
& + \frac{i\omega\tau}{6\pi\eta\xi}\frac{(4+3\epsilon^2)\hat a^3+36\epsilon^2\hat a^2+162\epsilon^2\hat a+81\epsilon^2}{2(9+9\hat a+\hat a^2)^2\epsilon^2},
\label{S:gamma_slltau}
\end{align}
where $\hat a=a/\xi$ and one can take further limits depending on the magnitude of $\hat a$.
In the high-frequency limit, $\omega \tau \to \infty$, $\mu$ becomes
\begin{align}
\mu[a,\omega]\approx \frac{1}{6\pi\eta a}-\frac{1}{6\pi\eta a (i\omega\tau)}.
\label{gamma_sggtau}
\end{align}
This expression is equivalent to writing the mobility as $\mu[a,\omega]\approx 1/(6\pi\eta_\mathrm{b} a)$, 
with an effective viscosity of $\eta_\mathrm{b}=\eta[1+1/(i\omega\tau)]$.
Therefore, the above two-fluid model reduces to the Kelvin-Voigt model at 
high-frequencies~\cite{LarsonBook}.

We previously obtained a general expression for the coupling mobility $M$ connecting the velocity 
$\mathbf{v}$ and the force $\mathbf{f}$ in the two-fluid model~\cite{Yasuda17b2,Yasuda17b},
which is also given in the SM.
For small distances, $r/\xi \to 0$, the coupling mobility is 
\begin{align}
M[r,\omega] \approx 
\frac{1}{4\pi\eta r}-\frac{1}{6\pi\eta \xi(1+i\omega \tau)^{1/2}}, 
\label{eq:Gshort}
\end{align}
and consequently, the gel is nearly purely viscous.
Conversely, for sufficiently large distances, $r/\xi  \to \infty$, the coupling mobility is
\begin{align}
M[r,\omega] & \approx 
\frac{1}{4\pi\eta_\mathrm{b} r}+\frac{\xi^2}{2\pi\eta r^3}\frac{1}{(1+i\omega\tau)^2}.
\label{eq:Glong}
\end{align}

\section{Average velocities in a two-fluid gel}

\begin{figure*}[tbh]
\begin{center}
\includegraphics[scale=0.28]{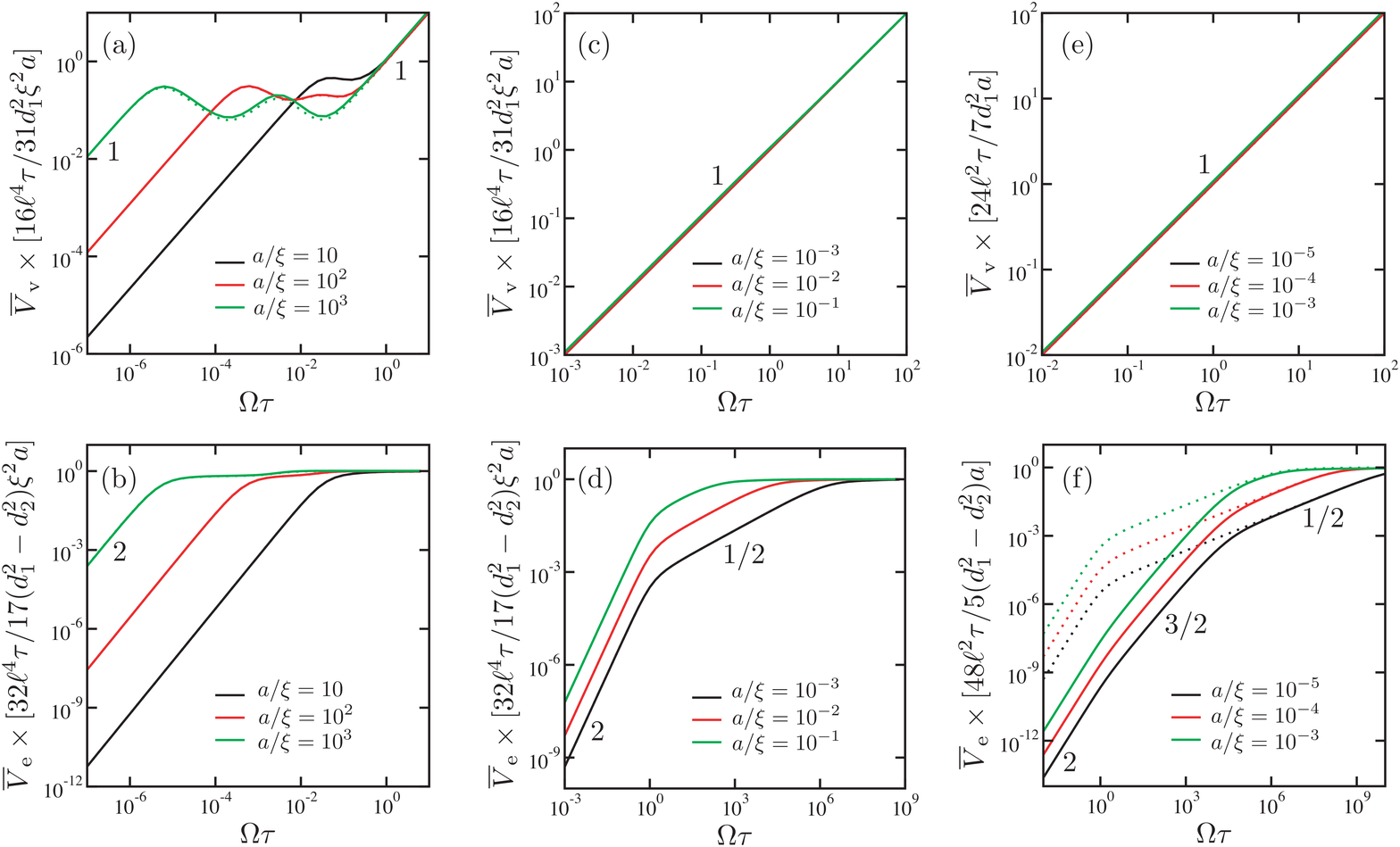}
\end{center}
\caption{
Plots of the average swimming velocity $\overline{V}$ as a function of the scaled frequency 
$\Omega \tau$ for a three-sphere micromachine swimming in a polymer gel with 
$\epsilon=[(K+4G/3)/G]^{1/2}=\sqrt{14/3}\approx 2.16$.
The solid lines are the numerical results explained in the text, whereas the dotted 
lines (with the same colors) are the analytical results obtained from eq.~(\ref{barVve}).
The analytical dotted curves are invisible when they coincide with the numerical solid curves.
The numbers indicate the slopes representing the exponents of the power-law behaviors.
(a) The scaled viscous contribution $\overline V_{\rm v}$ and (b) the scaled elastic 
contribution $\overline V_{\rm e}$ for a large swimmer ($a\gg\xi$, $\ell\gg\xi$).
The different colors correspond to the different $a/\xi$ values shown in the graphs; here, we have 
chosen $\ell/\xi=10^4$.
The other parameters are $\phi=\pi/2$ and $d_1/\ell=d_2/\ell=10^{-2}$  in (a) 
and $d_1=2 d_2$ and $d_2/\ell=10^{-2}$ in (b).
(c) The scaled $\overline V_{\rm v}$ and (d) scaled $\overline V_{\rm e}$ for a medium swimmer 
($a\ll\xi$, $\ell\gg\xi$).
The different colors correspond to the different $a/\xi$ values shown in the graphs; here, we have 
chosen $\ell/\xi=10^4$.
The other parameters are the same as those in (a) and (b).
(e) The scaled $\overline V_{\rm v}$ and (d) scaled $\overline V_{\rm e}$ for a small 
swimmer ($a\ll\xi$, $\ell\ll\xi$).
The different colors correspond to the different $a/\xi$ values shown in the graphs; here, we have
chosen $\ell/\xi=10^{-2}$.
The other parameters are the same as those in (a) and (b).
Notice that in all these plots, $a/\xi$ and $\ell/\xi$ are chosen in such a way that 
$a/\ell \ll 1$ is always satisfied.}
\label{plotV}
\end{figure*}

Next we discuss the average velocity of a three-sphere microswimmer in a two-fluid gel and investigate
its frequency as well as size dependencies. 
Recall that there are two lengths that measure the size of a three-sphere microswimmer:
the sphere radius $a$ and the average arm length $\ell$, with the condition $a \ll \ell$. 
Because the surrounding gel is characterized by the network mesh size, $\xi$, the following 
three different situations can be distinguished: 
(i) a large swimmer when $a\gg\xi$ and $\ell\gg\xi$, 
(ii) a medium swimmer when $a\ll\xi$ and $\ell\gg\xi$, and
(iii) a small swimmer when $a\ll\xi$ and $\ell\ll\xi$.
In each case, following the procedure in the SM, we numerically solve for $V_i$ and $F_i$ to 
calculate the average velocity $\overline V$ without making an expansion in terms of $a/\ell$.
These numerical results are plotted by the solid lines in fig.~\ref{plotV}, whereas 
the analytical results obtained from eq.~(\ref{barVve}) are plotted by the dotted lines
with the same colors.
In fig.~\ref{plotV}, the parameters $a/\xi$ and $\ell/\xi$ are chosen in such 
a way that the condition $a/\ell \ll 1$ is always satisfied for a three-sphere microswimmer.
Moreover, $d_1/\ell$ and $d_2/\ell$ are also small enough to ensure that 
the assumption of linear viscoelasticity is appropriate.

In figs.~\ref{plotV}(a) and (b), we separately plot the viscous $\overline V_{\rm v}$ and elastic
$\overline V_{\rm e}$ contributions, respectively, for a large swimmer as a function 
of the scaled frequency $\Omega \tau$.
The behavior of $\overline V_{\rm v}$ in fig.~\ref{plotV}(a) is remarkable because it exhibits a 
non-monotonic dependence on $\Omega$. 
A careful analysis reveals that it behaves as 
$\overline V_{\rm v} \sim\Omega \to \Omega^{-1/2} \to \Omega \to \Omega^{-1} \to \Omega$ 
as $\Omega$ increases.
This non-monotonic behavior is more pronounced for larger sphere sizes such as $a/\xi=100$ (green). 
On the other hand, the frequency dependence of the elastic contribution, $\overline V_{\rm e}$, crosses 
over as $\overline V_{\rm e} \sim\Omega^2\to\Omega^0$.

The above results can be reproduced by eq.~(\ref{barVve}) when we use the full expressions of the 
two-fluid mobilities $\mu$ and $M$ 
(most of the dotted lines in fig.~\ref{plotV} are invisible because they almost coincide with the solid lines).
Further, eq.~(\ref{barVve}) provides us with various asymptotic expressions.
For example, we first discuss the limit of $\Omega\tau\to0$  in $\overline V_{\rm v}$.
Using the first term of eq.~(\ref{S:gamma_slltau}) and the second term of eq.~(\ref{eq:Glong}), 
we obtain 
\begin{align}
\overline V_{\rm v} \approx\frac{31d_1d_2a^3\Omega}{144\ell^4}\sin\phi, 
\label{V1}
\end{align}
which is proportional to $\Omega$.
The complex non-monotonic behaviors in the intermediate frequencies are separately discussed in the SM. 
In the limit of $\Omega\tau\to\infty$, we use the first term of eq.~(\ref{gamma_sggtau}) and 
the second term of eq.~(\ref{eq:Glong}) to obtain 
\begin{align}
\overline V_{\rm v} \approx\frac{31d_1d_2\xi^2a\Omega}{16\ell^4}\sin\phi. 
\label{LS_VH}
\end{align}

By evaluating the elastic term in eq.~(\ref{barVve}), one can also obtain the asymptotic expressions 
for $\overline V_{\rm e}$.
In the limit of $\Omega\tau\to0$, we use the second terms of eqs.~(\ref{S:gamma_slltau}) and (\ref{eq:Glong}) 
to obtain 
\begin{align}
\overline V_{\rm e} \approx\frac{17(d_1^2-d_2^2)a^5}{5184\ell^4\xi^2\tau}\frac{4+3\epsilon^2}{\epsilon^2}(\Omega\tau)^2,
\label{V2}
\end{align}
which is proportional to $\Omega^2$.
In the limit of $\Omega\tau\to\infty$, the second terms of eqs.~(\ref{gamma_sggtau}) and 
(\ref{eq:Glong}) yield 
\begin{align}
\overline V_{\rm e} \approx\frac{17(d_1^2-d_2^2)\xi^2a}{32\ell^4\tau},
\label{LS_EH}
\end{align}
which is independent of $\Omega$.
Notice that the crossover frequency separating the different scaling regimes 
is strongly dependent on $a/\xi$. 
For example, the crossover frequency between eqs.~(\ref{V2}) and (\ref{LS_EH}) roughly scales 
as $(\Omega\tau)^{\ast} \sim (a/\xi)^{-2}$. 
This means that one can extract information concerning the internal structure of the surrounding medium 
by observing the average swimming velocity.
Moreover, the non-linear dependence on the sphere size, such as $\overline V_{\rm v} \sim a^3$ 
in eq.~(\ref{V1}) or $\overline V_{\rm e} \sim a^5$ in eq.~(\ref{V2}), is also a unique feature of a 
two-fluid gel.

For a medium swimmer ($a\ll\xi$ and $\ell\gg\xi$), we plot the numerical results of 
$\overline V_{\rm v}$ and $\overline V_{\rm e}$ in figs.~\ref{plotV}(c) and (d), respectively.
In fig.~\ref{plotV}(c), we find $\overline V_{\rm v} \sim\Omega$ over the entire frequency range.
This behavior is essentially explained by eq.~(\ref{LS_VH}), which was obtained for the large swimmer
case.
The elastic contribution in fig.~\ref{plotV}(d), conversely, 
crosses over as $\overline V_{\rm e} \sim\Omega^2 \to \Omega^{1/2}\to\Omega^0$, 
where the $\Omega$-independent behavior can be explained by eq.~(\ref{LS_EH}) as before.
In the SM, we show the asymptotic expressions for smaller frequencies (see eqs.~(S40) and (S41)).

Finally, we numerically plot $\overline V_{\rm v}$ and $\overline V_{\rm e}$ for a small swimmer 
($a\ll\xi$ and $\ell\ll\xi$) in figs.~\ref{plotV}(e) and (f), respectively.
The viscous contribution shows a linear dependence, $\overline V_{\rm v} \sim\Omega$.
This is reasonable because the combination of the first terms in eqs.~(\ref{gamma_sggtau}) and 
(\ref{eq:Gshort}) simply represents a purely viscous fluid. 
The elastic contribution plotted in fig.~\ref{plotV}(f) crosses over as  
$\overline V_{\rm e} \sim \Omega^2 \to \Omega^{3/2} \to \Omega^{1/2} \to \Omega^0$.
Even though the first two scaling behaviors cannot be obtained analytically, the last two behaviors 
are given in the SM (see eqs.~(S42) and (S43)).
It should be mentioned here that eq.~(\ref{barVve}) does not reproduce the numerical result in 
fig.~\ref{plotV}(f) because the lowest expansion in terms of $a/\ell$ is inappropriate 
for this limit.
Hence the dotted lines deviate from the solid lines fig.~\ref{plotV}(f) especially for 
relatively smaller $\Omega \tau$ values.

\section{Discussion}

So far, we have primarily discussed the motion of a three-sphere microswimmer in a polymer gel 
described by the two-fluid model.
However, the importance of our work is not restricted to these specific models.
The prediction of the average velocity in eq.~(\ref{barVve}) is applicable to any structured 
viscoelastic fluid that has an intermediate length scale and a characteristic time scale. 
For example, one can also discuss the motion of a microswimmer in a polymer solution 
that is described by a different ``two-fluid" model~\cite{Doi92,Bruinsma14}. 
In addition, it is interesting to discuss the dynamics of three-sphere microswimmers in liquid 
crystals~\cite{Krieger15}, which typically exhibit complex rheological behavior depending on 
their different phases~\cite{Fujii14}.

Furthermore, we encounter a similar situation when we consider the motion of a three-disk 
microswimmer immersed in a quasi-2D fluid membrane~\cite{Ota18}. 
Owing to the presence of the hydrodynamic screening length in the quasi-2D fluid, 
the geometric factor appearing in the average velocity exhibits various asymptotic behaviors 
when changing the ratio between the swimmer size and the screening length.
The result in ref.~\cite{Ota18} can be obtained from eq.~(\ref{barVve}) using the 
mobility of the disk and the coupling mobility in a quasi-2D fluid.

At this point, it is useful to give some numbers related to realistic microswimmers and 
systems described by a two-fluid model.
Let us first consider a homogeneous sample of entangled F-actin network whose mesh size
$\xi$ can be controlled by the actin monomer concentration $c$ as 
$\xi = 0.3/\sqrt{c}$ ($\xi$ in $\mu$m and $c$ in mg/ml)~\cite{Sonn14,Sonn14-b}.
Hence $\xi \approx 1$~$\mu$m for $c\approx 0.1$~mg/ml.
On the other hand, a typical size of colloidal particles used in microrheology experiments 
is $a \approx 0.5$~$\mu$m~\cite{FurstBook}.
If we assume that one can construct a three-sphere microswimmer by connecting these colloidal
particles, its whole size (corresponding to $\ell$) would amount to several microns, which is  
comparable to a typical size of living microorganisms such as bacteria.
Since a magnitude relation $a<\xi<\ell$ holds in this case, a living microorganism swimming 
in an actin network may correspond to a medium swimmer (see fig.~\ref{barVve}(c) and (d)).

On the other hand, a three-sphere swimmer has  been experimentally realized 
by using ferromagnetic particles at an air-water interface and by applying an oscillating magnetic field~\cite{Grosjean16,Grosjean18}.
In this experiment, the size of spheres is about $a\approx 200$~$\mu$m and the 
rest length of the bonds is roughly $\ell \approx 1$~mm.
Hence such a model swimmer behaves as a large swimmer in an actin network (see 
fig.~\ref{barVve}(a) and (b)), since the mesh size is much smaller, i.e.,  $\xi \ll a<\ell$.
Such a comparison shows that the swimming behaviors of a microorganism and a ferromagnetic 
swimmer are essentially different in a structured fluid such as F-actin networks due to the presence 
of the characteristic length scale $\xi$. 
This is the main message of the present work.

Concerning the characteristic time scale $\tau$ of a two-fluid model, we shall 
refer to a number that was measured for F-actin networks~\cite{Sonn14,Sonn14-b} by using 
microrheology techniques~\cite{FurstBook}.
With the water viscosity $\eta \approx 10^{-3}$~Pa$\cdot$s and the measured shear 
modulus $G \approx 10^{-1}$~Pa for a F-actin network, we obtain the viscoelastic time scale as 
$\tau = \eta/G \approx 10^{-2}$~s.
For a typical microorganism and a model swimmer mentioned above, characteristic frequencies 
are about $\Omega \approx 10^2$~Hz~\cite{Lauga09a} and $1$~Hz~\cite{Grosjean16,Grosjean18}, 
respectively.
Hence the corresponding Deborah numbers for these two cases are ${\rm De} \approx 1$ 
(microorganism) and $10^{-2}$ (model swimmer) which are both considered in fig.~\ref{barVve}.

Finally, we note that Fu \textit{et al.}\ analyzed the swimming behavior of an infinite sheet undergoing 
transverse traveling-wave deformations in a two-fluid gel~\cite{Fu10}.
They demonstrated that the boundary conditions between the sheet and the network significantly affect the 
swimming speed.
In our study, we considered only the case of ``a sticking fluid and a free network" for the 
boundary condition between the sphere and the gel. 
Different situations such as ``a sticking fluid and a sticking network" case or  
``a sticking fluid and a slipping network" case, as discussed in detail in ref.~\cite{Diamant15}, will 
lead to different swimming behaviors because the self-mobility of the spheres is modified.
Although several deficiencies of the two-fluid model have been explicitly pointed 
out~\cite{Diamant15}, such investigations are left to future studies.

\section{Summary}

We discussed the locomotion of a three-sphere microswimmer in a viscoelastic 
structured fluid with typical length and time scales. 
We derived a general expression for the average swimming velocity,  eq.~(\ref{barVve}),
which includes both viscous and elastic contributions. 
To illustrate our result, we used the two-fluid model for a polymer gel and demonstrated that 
the average velocity exhibits various asymptotic behaviors depending on the swimmer size.
Because one can extract information concerning the internal structure of the surrounding fluid 
from observations of the motion of a microswimmer, the present theory offers a new 
approach to active microrheology~\cite{FurstBook}.

\acknowledgments

K.Y.\ acknowledges support by a Grant-in-Aid for JSPS Fellows　(Grant No.\ 18J21231)　from the Japan 
Society for the Promotion of Science (JSPS).
S.K.\ acknowledges support by a Grant-in-Aid for Scientific Research (C) (Grant No.\ 18K03567) 
from the JSPS.


\end{document}